\documentclass[12pt]{article}
  \usepackage{amsfonts}
  \usepackage{amsmath}
\usepackage{amssymb}
\usepackage{amscd}
\usepackage[dvips]{graphicx}
\begin{document}
~~
\bigskip
\bigskip
\begin{center}
{\Large {\bf{{{Newton equation for canonical, Lie-algebraic and
quadratic deformation of classical space}}}}}
\end{center}
\bigskip
\bigskip
\bigskip
\begin{center}
{{\large ${\rm {Marcin\;Daszkiewicz^{1}}}$, \large ${\rm {Cezary\;
 J.\;Walczyk^{2}}}$ }}
\end{center}
\bigskip
\begin{center}
\bigskip

{
${\rm{~^{1}Institute\; of\; Theoretical\; Physics}}$}

{
${\rm{ University\; of\; Wroclaw\; pl.\; Maxa\; Borna\; 9,\;
50-206\; Wroclaw,\; Poland}}$}

{
${\rm{ e-mail:\; marcin@ift.uni.wroc.pl}}$}

\bigskip

{
${\rm{~^{2}Department\; of\; Physics}}$}

{
${\rm{ University\; of\; Bialystok,\; ul.\; Lipowa\; 41,\;
15-424\;Bialystok,\; Poland}}$}

{
${\rm{ e-mail:\; c.walczyk@alpha.uwb.edu.pl}}$}

\end{center}
\bigskip
\bigskip
\bigskip
\bigskip
\bigskip
\bigskip
\bigskip
\bigskip
\bigskip
\begin{abstract}
The Newton equation describing  the particle motion in  constant
external field force on canonical, Lie-algebraic and quadratic
space-time is investigated. We show that  for canonical  deformation
of space-time the dynamical effects are absent, while in the case of
Lie-algebraic noncommutativity, when spatial coordinates  commute to
the time variable, the additional acceleration of particle is
generated. We also indicate, that in the case of spatial coordinates
commuting in Lie-algebraic way, as well as for quadratic
deformation, there appear additional velocity and position-dependent
forces.
\end{abstract}
\bigskip
\bigskip
\bigskip
\bigskip
\eject

\section{{{Introduction}}}

~~~\,\,\,Due to several  theoretical arguments (see e.g.
\cite{2}-\cite{2s}) the interest in  studying of space-time
noncommutativity is growing rapidly. There appeared a lot of papers
dealing with noncommutative classical (\cite{romero}-\cite{kochan})
and quantum (\cite{qm0}-\cite{lodzianieosc}) mechanics, as well as
with field theoretical models (see e.g.
\cite{fieldus}-\cite{maria}), defined on quantum
 space-time.

At present, in accordance with the Hopf-algebraic classification of
all deformations of relativistic and nonrelativistic symmetries
\cite{zakrzewski}, \cite{kowclas}, one can distinguish three kinds
of
space-time noncommutativity:\\
\\
{ 1)} Canonical (soft) deformation
\begin{equation}
[\;{ x}_{\mu},{ x}_{\nu}\;] = i\theta_{\mu\nu}\;, \label{noncomm}
\end{equation}
with tensor $\theta_{\mu\nu}$ being constant and antisymmetric
($\theta_{\mu\nu} = -\theta_{\nu\mu}$). The explicit form of
corresponding Poincare Hopf algebra has been provided in \cite{3a},
while  its nonrelativistic  counterpart has been
proposed in \cite{3b}. \\
\\
{ 2)} Lie-algebraic  case 
\begin{equation}
[\;{ x}_{\mu},{ x}_{\nu}\;] = i\theta_{\mu\nu}^{\rho}x_{\rho}\;,
\label{noncomm1}
\end{equation}
with  particularly chosen constant coefficients
$\theta_{\mu\nu}^{\rho}$.
This kind of space-time modification is represented by
$\kappa$-Poincare \cite{4a}, \cite{4b} and $\kappa$-Galilei
\cite{kgalilei} Hopf algebras. Besides, the Lie-algebraic twist
deformations of relativistic and nonrelativistic symmetries have
been provided in \cite{lie2}, \cite{lie1} and
\cite{3b}\footnote{There also exist so-called fuzzy space
noncommutativity \cite{fuzzy}. However, in this article such a type
of deformation will be not under consideration.}.
\\
\\
{ 3)} Quadratic deformation
\begin{equation}
[\;{ x}_{\mu},{ x}_{\nu}\;] =
i\theta_{\mu\nu}^{\rho\tau}x_{\rho}x_{\tau}\;, \label{noncomm2}
\end{equation}
with constant coefficients $\theta_{\mu\nu}^{\rho\tau}$. Its
Hopf-algebraic realization was proposed in \cite{lie2}, \cite{slawa}
in the case of relativistic symmetry, and in \cite{3c}, for its
nonrelativistic counterpart.

 In this
article we investigate the impact of the mentioned above
nonrelativistic deformations (with commuting time
direction)\footnote{We consider
 only spatial deformations, i.e. time plays a role of parameter.} 
on  dynamics of simplest classical  system - the nonrelativistic
particle moving in a field of constant force. We indicate that in
the case of soft
deformation the Newton equation is not modified, while 
for the Lie-algebraic  noncommutativity  we recover two interesting
dynamical effects. First of them corresponds to  the
case, when commutator of two spatial directions closes to time coordinate, 
and
then, 
such a kind of noncommutativity   additionally produces the
acceleration of moving particle. For the second type of
Lie-algebraic deformation, when the commutator of spatial directions
closes to space coordinates, 
there are generated  
  velocity and position-dependent forces, i.e. forces,
 which
depend on velocity $( \dot{x})$ and position $( {x})$ of moving
particle, respectively\footnote{ In the case of "position force" one
can recognize  well-known inverted oscillator force (see e.g.
\cite{invosc}).}.

In the case of quadratic  deformation  the situation appears  most
complicated. Similarly  to the
Lie-algebraic case, this type of noncommutativity generates new
velocity as well as  position-dependent forces, but this time, with
an explicit
time-dependence. 
In this paper the analytic form of the corresponding solutions is
presented and analyzed  in detail.

The paper is organized as follows. In  first Section we review some
known facts concerning the classical mechanics on canonically
deformed  quantum space (see e.g \cite{romero}). We
indicate that in such a case the Newton equation for particle moving
in a constant force remains unchanged. In Section 2 we analyze two
cases of Lie-algebraic  deformations, and we provide the
corresponding
phase spaces as well as we solve suitable Newton equations. 
Section 3 deals with the quadratic deformation of classical space.
The corresponding  Newton equation is provided and its solution is
studied as well. The results are summarized and discussed in the
last Section.

\section{{{Canonical noncommutativity}}}

Let us start with a set of variables $\zeta^a$ with $a = 1,2,\ldots
,2n$. For arbitrary two functions $F(\zeta^a)$ and $G(\zeta^a)$ we
 define  Poisson bracket as follows (\cite{sym}; for application to
 noncommutative space-time see \cite{mech1})
\begin{equation}
\{\,F,G\,\}=\{\,\zeta^a,\zeta^b\,\}\ \frac{\partial F}{\partial
\zeta^a}\ \frac{\partial G}{\partial \zeta^b}\;.\label{symp}
\end{equation}
In terms of the above structure and given  Hamiltonian $H =
H(\zeta^a)$ one can write the equations of motion as
\begin{equation}
\dot{\zeta}^a=\{\,\zeta^a,H\,\}\;.\label{motion2}
\end{equation}
In general case (for any function $F$ depending on $\zeta^a$) we
have
\begin{equation}
\dot{F}=\{\,F,H\,\}\;.\label{motion1}
\end{equation}
Below, we will consider the phase space given by $\zeta^a =
(x_i,p_i)$ with $i=1,2,3$.

Let us   start with   canonical type of  noncommutativity
\begin{equation}
\{\,x_i,x_j\,\} = \theta_{ij}\;,\label{soft1}
\end{equation}
supplemented   by
\begin{equation}
\{\,p_i,p_j\,\}=0\;\;\;,\;\;\;\{\,x_i,p_j\,\}=\delta_{ij}\;.
\label{can}
\end{equation}
The relations (\ref{soft1}) and (\ref{can}) define the symplectic
structure for the soft deformation of classical (commutative) space,
which was studied in
\cite{romero}-\cite{aca}. \\
In accordance with (\ref{motion2}) for the Hamiltonian
\begin{eqnarray}
H(\vec{p},\vec{x}) = \frac{\vec{p}^{\,2}}{2m} +
V(\vec{x})\;,\label{ham}
\end{eqnarray}
we get the following equations of motion
\begin{eqnarray}
\dot{x}_i &=&  \theta_{ik}\,\frac{\partial V}{\partial x_k} +
\frac{p_i}{m}\;\;\;,\;\;\;
 \dot{p}_i =  -\frac{\partial V}{\partial
x_i}\;. \label{ham2}
\end{eqnarray}
They lead to the corresponding  Newton equation (see e.g.
\cite{romero})
\begin{eqnarray}
m\ddot{x}_i  = m \theta_{ik}\,\frac{d}{dt}\left(\frac{\partial
V}{\partial x_k}\right)-\frac{\partial V}{\partial x_i} = m
\theta_{ik}\,\frac{\partial^2 V}{\partial x_k \partial
x_l}\dot{x}_l-\frac{\partial V}{\partial x^i} \;, \label{cannewton}
\end{eqnarray}
which  for   potential\footnote{In the case of quantum space the
differential calculus  is highly-nontrivial (see e.g. \cite{oeckl},
\cite{sitarz}). Fortunately, for such a simple linear  function like
the potential (\ref{potential}) (there is no products of spatial
variables) the result of differentiation  is classical.}
\begin{eqnarray}
V(\vec{x}) = -\sum_{i=1}^{3}F_{i}x_{i}\;\;\;,\;\;\;\frac{\partial
V}{\partial x_k} = -F_k = {\rm const.}\;, \label{potential}
\end{eqnarray}
 remains not deformed
\begin{eqnarray}
m\ddot{x}_i  = -\frac{\partial V}{\partial x_i} = F_i\;.
\label{unnewton}
\end{eqnarray}
Hence, we see that the canonical space-time deformation
(\ref{soft1}) does not provide any  dynamical effects for particle
moving in the potential (\ref{potential}) corresponding to constant
force.

\section{{{Lie-algebraic noncommutativity}}}

\subsection{{{Space coordinates commuting  to time}}}

Let us consider the  Lie-algebraic deformation of space with
two spatial directions commuting to time in the following way 
\begin{eqnarray}
\{\,x_i,x_j\,\} = \frac{1}{\kappa} t(\delta_{i\rho}\delta_{j\tau} -
\delta_{i\tau}\delta_{j\rho}) \;, \label{lie1}
\end{eqnarray}
where $\kappa$ is the mass-like
 deformation parameter;  indices $\rho$, $\tau$ are
different and fixed. As it was already mentioned, such a type of
noncommutativity has been recovered in a Hopf algebraic framework in
\cite{kowclas}, \cite{3b}  with use of the contraction procedure
\cite{cont1}, \cite{cont2}. Its relativistic counterpart has been
proposed
 in  \cite{lie2}.  \\
The  commutation relations (\ref{lie1}) can be   extended (in
accordance with Jacobi identity) to the whole phase space as follows
\begin{eqnarray}
\{\,p_i,p_j\,\}=0\;\;\;,\;\;\;\{\,x_i,p_j\,\}=\delta_{ij}\;.
\label{ssslie1}
\end{eqnarray}
In such a case the Hamilton
 equations (\ref{motion2})
 take the form
\begin{eqnarray}
\dot{x}_i &=&  t(\delta_{i\rho}\delta_{k\tau} -
\delta_{i\tau}\delta_{k\rho})\,\frac{1}{\kappa}\frac{\partial
V}{\partial x_k}
 +
\frac{p_i}{m}\;\;\;,\;\;\;
 \dot{p}_i =  -\frac{\partial V}{\partial
x_i}\;, \label{lieham21}
\end{eqnarray}
while the corresponding Newton equation looks as follows
\begin{eqnarray}
m\ddot{x}_i  = t(\delta_{i\rho}\delta_{k\tau} -
\delta_{i\tau}\delta_{k\rho})\,\frac{m}{\kappa}\frac{d}{dt}\left(\frac{\partial
V}{\partial x_k}\right)  + (\delta_{i\rho}\delta_{k\tau} -
\delta_{i\tau}\delta_{k\rho})\,\frac{m}{\kappa}\frac{\partial
V}{\partial x_k} -\frac{\partial V}{\partial x_i}\;.
\label{lie1newton}
\end{eqnarray}
For the simplest potential (\ref{potential}) we have\footnote{Due to
the fact, that in the equation (\ref{lie1newton1}) there is no
product of two spatial (noncommutative) positions and velocities,
the considering equation is represented on commutative space by the
formula (\ref{lie1newton1}) as well. In other words, we can pass
with Newton equation (\ref{lie1newton1}) to the undeformed space
without using any  star product \cite{3b} (a Weyl map \cite{bloch}).
The same situation appears as well in the case of others considered
deformations.}

\begin{equation}
\left\{\begin{array}{rcl} m\ddot{x}_i  &=&  F_i\;
\\&~~&~\cr
m\ddot{x}_\rho &=& -\dfrac{m}{\kappa} F_\tau +F_\rho\;\\&~~&~\cr
m\ddot{x}_\tau &=& \dfrac{m}{\kappa} F_\rho
+F_\tau\;,\end{array}\right.\label{lie1newton1}
\end{equation}
\\
with index $i$ different from $\rho$ and $\tau$. \\
By trivial integration one can find the following solution  of the
above  system

\begin{eqnarray}
{x}_i (t) &=&  \frac{F_i}{2m}t^2 + v_{i0}t +
x_{i0}\;\;\;,\;\;\;{x}_\tau (t)=\frac{(\frac{m}{\kappa} F_\rho
+F_\tau)}{2m}t^2+v_{\tau 0}t + x_{\tau 0} \;,\label{solu1}
\\&~~&~\cr
&&~~~~~~~{x}_\rho(t) = \frac{(-\frac{m}{\kappa} F_\tau
+F_\rho)}{2m}t^2
+v_{\rho 0}t + x_{\rho 0} \;,\label{solu2} 
\end{eqnarray}
\\
where $x_{a0}$  and $v_{a0}$  $(a = k,l)$ denote   initial positions
and velocities, respectively.

We see, that the  noncommutativity (\ref{lie1}) generates additional
acceleration of particle in  fixed directions $\rho$ and $\tau$. In
direction $i$ the motion of particle remains undeformed. Of course,
for $\kappa \to \infty$, the above solutions become classical and
describe   particle moving in external constant force
$\vec{F}=[\,F_i,F_\tau,F_\rho\,]$.

\subsection{{{Space coordinates commuting to space}}}

Let us now turn to the case when two spatial directions commute to
the spatial ones\footnote{$\left[\,{\hat \kappa}\,\right] = {\rm
N}\cdot {\rm s}$.}
\begin{equation}
\{\,x_k,x_\gamma\,\} = \frac{1}{\hat \kappa} x_l\;\;\;,\;\;\;
 \{\,x_l,x_\gamma\,\} = -\frac{1}{\hat \kappa} x_k\;\;\;,\;\;\;\{\,x_k,x_l\,\} = 0\;,\label{hlie2a}
\end{equation}
and where   indices $k$, $l$, $\gamma$ are different and fixed. Such
a type of noncommutativity  has been proposed in the case of
nonrelativistic symmetry in \cite{kowclas} as the translation sector
of classical Poisson-Lie structure, and in \cite{3b},  as the Hopf
module of  quantum Galilei   algebra. Its relativistic
counterpart has been obtained in  \cite{lie2}. \\
The corresponding phase space is given by the  Poisson brackets
(\ref{hlie2a}) augmented  by
\begin{eqnarray}
&&\{\,p_k,x_\gamma\,\} = \frac{1}{\hat \kappa}
p_l\;,\;\{\,p_l,x_\gamma\,\} = -\frac{1}{\hat \kappa}
p_k\;,\;\{\,x_i,p_j\,\}
=\delta_{ij}\;,\cr
&~~&~\cr
&&~~~~~~~~~~~~~~~\{\,x_\gamma,p_\gamma\,\}=1\;,\;\{\,p_a,p_b\,\}=0
 \;,\nonumber
\end{eqnarray}
where indices $i,j$ are different from $\gamma$ and $a,b = 1,2,3$.\\
Using  the formula (\ref{motion2}) one can find the following
 equations of motion
\begin{equation}
\dot{x}_{k} \;=\;  \frac{p_{k}}{m} \;+\; x_{l}\,\frac{1}{\hat
\kappa}\frac{\partial V}{\partial x_\gamma}\;\;\;,\;\;\;
 \dot{p}_{k} \;=\;  p_{l}\,\frac{1}{\hat \kappa}\frac{\partial V}{\partial
x_\gamma}\;-\;\frac{\partial V}{\partial x_{k}}\;,~~ \label{zadrug0}
\end{equation}
\begin{equation}
\dot{x}_{l} \;=\; \frac{p_{l}}{m} \;-\; x_{k}\,\frac{1}{\hat
\kappa}\frac{\partial V}{\partial x_\gamma}\;\;\;,\;\;\; \dot{p}_{l}
\;=\; -\frac{\partial V}{\partial x_{l}}\; -\; p_{k}\,\frac{1}{\hat
\kappa}\frac{\partial V}{\partial x_\gamma}\;, \label{zadrug2}
\end{equation}
\\
in  $k$, $l$-directions, and
\begin{eqnarray}
\dot{x}_{\gamma} &=&  \frac{p_{\gamma}}{m} - x_{l}\,\frac{1}{\hat
\kappa}\frac{\partial V}{\partial x_k}  + x_{k}\,\frac{1}{\hat
\kappa}\frac{\partial V}{\partial x_l}\;,\\&~~&~\cr
&&~~~\dot{p}_{\gamma} \;=\; -\frac{\partial V}{\partial
x_{\gamma}}\;,\label{zadrug4}
\end{eqnarray}
in  $\gamma$-direction. \\
The corresponding Newton equations can be find  with use of
(\ref{zadrug0})-(\ref{zadrug4})

\begin{eqnarray}
m\ddot{x}_k  &=&  -\frac{\partial V}{\partial
x_k}+x_l\,\frac{m}{\hat \kappa}\frac{d}{dt}\left(\frac{\partial
V}{\partial x_\gamma}\right) +\dot{x}_l\,\frac{2m}{\hat
\kappa}\frac{\partial
V}{\partial x_\gamma} +\nonumber\\
&~~&~\cr &&~~~~~~~~+mx_k\left(\frac{1}{\hat
\kappa}\right)^2\left(\frac{\partial V}{\partial
x_\gamma}\right)^2\;,
\label{lie2newton0}\\
&~~&~\cr m\ddot{x}_l  &=&  -\frac{\partial V}{\partial
x_l}-x_k\,\frac{m}{\hat \kappa}\frac{d}{dt}\left(\frac{\partial
V}{\partial x_\gamma}\right) -\dot{x}_k\,\frac{2m}{\hat
\kappa}\frac{\partial V}{\partial x_\gamma}+\nonumber\\
&~~&~\cr &&~~~~~~~~+mx_l\left(\frac{1}{\hat
\kappa}\right)^2\left(\frac{\partial V}{\partial
x_\gamma}\right)^2\;,
\label{lie2newton00}\\
&~~&~\cr m\ddot{x}_\gamma  &=&  -\frac{\partial V}{\partial
x_\gamma} -x_l\,\frac{m}{\hat
\kappa}\frac{d}{dt}\left(\frac{\partial V}{\partial x_k}\right)
-\dot{x}_l\,\frac{m}{\hat \kappa}\frac{\partial
V}{\partial x_k} +\nonumber\\
&~~&~\cr &&~~~~~~~~~~~~~~+
 x_k\,\frac{m}{\hat \kappa}\frac{d}{dt}\left(\frac{\partial V}{\partial
 x_l}\right)+
\dot{x}_k\,\frac{m}{\hat \kappa}\frac{\partial V}{\partial x_l}
 \;,
\label{lie2newton1}
\end{eqnarray}
\\
and for the potential (\ref{potential}) they look as follows 

\begin{equation}
\left\{\begin{array}{rcl} m\ddot{x}_k  &=&  F_k -\dfrac{2m}{\hat
\kappa}F_\gamma\dot{x}_l +m\left(\dfrac{F_\gamma}{\hat
\kappa}\right)^2x_k\;\\
&~~&~\cr m\ddot{x}_l  &=& F_l+\dfrac{2m}{\hat
\kappa}F_\gamma\dot{x}_k +m\left(\dfrac{F_\gamma}{\hat
\kappa}\right)^2x_l\;\\
&~~&~\cr m\ddot{x}_\gamma  &=&  F_\gamma +\dfrac{m}{\hat
\kappa}F_k\dot{x}_l - \dfrac{m}{\hat \kappa}F_l\dot{x}_k
 \;.
\label{slie2newton0}\end{array}\right.
\end{equation}
\\
$~~~~~$We see, that this kind of space-time deformation generates
 velocity ($F \sim \dot{x}$) and position-dependent
($F \sim {x}$) forces
 corresponding to  both directions $k$ and $l$. As it was mentioned in Introduction
 (see footnote 3), in the position dependent force
 we recognize  well-known inverted oscillator force
 \cite{invosc}. Besides,
 by direct calculations one can also check that
the solution of above system is given by formulae

\begin{eqnarray}
{x}_k(t)  &=&  -\frac{F_k{\hat \kappa}^2}{F_\gamma^2m}
+\left[t\left(\frac{F_l{\hat \kappa}}{F_\gamma m}+v_{k0}+
\frac{F_\gamma x_{l0}}{{\hat \kappa}}\right)+x_{k0} + \frac{F_k{\hat
\kappa}^2}{F_\gamma^2m} \right]\cdot \cos \left(\frac{F_\gamma
t}{{\hat
\kappa}}\right)+\nonumber\\
&~~&~\cr &&~~~~~~~- \left[t\left(\frac{F_k{\hat \kappa}}{F_\gamma
m}-v_{l0}+ \frac{F_\gamma x_{k0}}{{\hat \kappa}}\right)-x_{l0} +
\frac{F_l{\hat \kappa}^2}{F_\gamma^2m} \right]\cdot \sin
\left(\frac{F_\gamma t}{{\hat \kappa}}\right)\;,\label{sso0}
\\
&~~&~\cr {x}_l(t)  &=& -\frac{F_l{\hat \kappa}^2}{F_\gamma^2m}
+\left[t\left(-\frac{F_k{\hat \kappa}}{F_\gamma m}+v_{l0}-
\frac{F_\gamma x_{k0}}{{\hat \kappa}}\right)+x_{l0} + \frac{F_l{\hat
\kappa}^2}{F_\gamma^2m} \right]\cdot \cos \left(\frac{F_\gamma
t}{{\hat
\kappa}}\right)+\nonumber\\
&~~&~\cr &&~~~~~~~+ \left[t\left(\frac{F_l{\hat \kappa}}{F_\gamma
m}+v_{k0}+ \frac{F_\gamma x_{l0}}{{\hat \kappa}}\right)+x_{k0} +
\frac{F_k{\hat \kappa}^2}{F_\gamma^2m} \right]\cdot \sin
\left(\frac{F_\gamma t}{{\hat \kappa}}\right)\;,\\
&~~&~\cr {x}_\gamma(t)  &=&\frac{F_\gamma}{2m}t^2 + v_{0\gamma}t +
x_{0\gamma}+ t\frac{1}{{\hat
\kappa}}(F_lx_{k0}-F_kx_{l0})+\nonumber\\
&~~&~\cr &+&
\frac{1}{F_\gamma}\left[\,\frac{1}{F_\gamma}\left(\frac{2F_k^2{\hat
\kappa}^2}{F_\gamma m} +\frac{2F_l^2{\hat \kappa}^2}{F_\gamma m} +
{F_l{\hat \kappa}v_{k0}}- {F_k{\hat \kappa}v_{l0}}\right) +
{2F_kx_{k0}}+ {2F_lx_{l0}}\,\right] +\nonumber\\
&~~&~\cr &+& \left[ t\left(\frac{1}{F_\gamma}\left(\frac{F_k^2{\hat
\kappa}}{F_\gamma m} + \frac{F_l^2{\hat \kappa}}{F_\gamma m} +
{F_lv_{k0}} - {F_kv_{l0}}\right)+ \frac{F_kx_{k0}}{{\hat \kappa}} +
\frac{F_lx_{l0}}{{\hat \kappa}}\right)
- \frac{F_k{\hat \kappa}v_{k0}}{F_\gamma^2}\,+\right. \nonumber\\
&~~&~\cr &&~~~~~~~~~~~~~~-\left. \frac{F_l{\hat
\kappa}v_{l0}}{F_\gamma^2}+\frac{2}{{F_\gamma}}(F_lx_{k0} -
{F_kx_{l0}})\right]\cdot \sin \left(\frac{F_\gamma t}{{\hat
\kappa}}\right)+ \label{ssol}\\
&~~&~\cr &&~~~~~~~- \left[
t\left(\frac{1}{F_\gamma}\left({F_kv_{k0}} + {F_lv_{l0}}\right)  -
\frac{F_lx_{k0}}{{\hat \kappa}}+
\frac{F_kx_{l0}}{{\hat \kappa}}\right)
+ \frac{2F_k^2{\hat \kappa}^2}{F_\gamma^3m}\,+\right. \nonumber\\
&~~&~\cr &+&\left. \frac{1}{F_\gamma}\left( \frac{2F_l^2{\hat
\kappa}^2}{F_\gamma^2m}+\frac{F_l{\hat \kappa}v_{k0}}{F_\gamma}-
\frac{F_k{\hat \kappa}v_{l0}}{F_\gamma} + 2({F_k x_{k0}} +
{F_lx_{l0}}) \right)\right]\cdot \cos \left(\frac{F_\gamma t}{{\hat
\kappa}}\right)\;,\nonumber
\end{eqnarray}
\\
where $x_{a0}$ and $v_{a0}$ $(a = k,l)$ denote   initial positions
and velocities, respectively. The corresponding trajectories are
illustrated  on Figure 1 for
different values of parameter ${\hat \kappa}$. 
\begin{figure}[!h]
\begin{center}
\includegraphics[width=14cm]{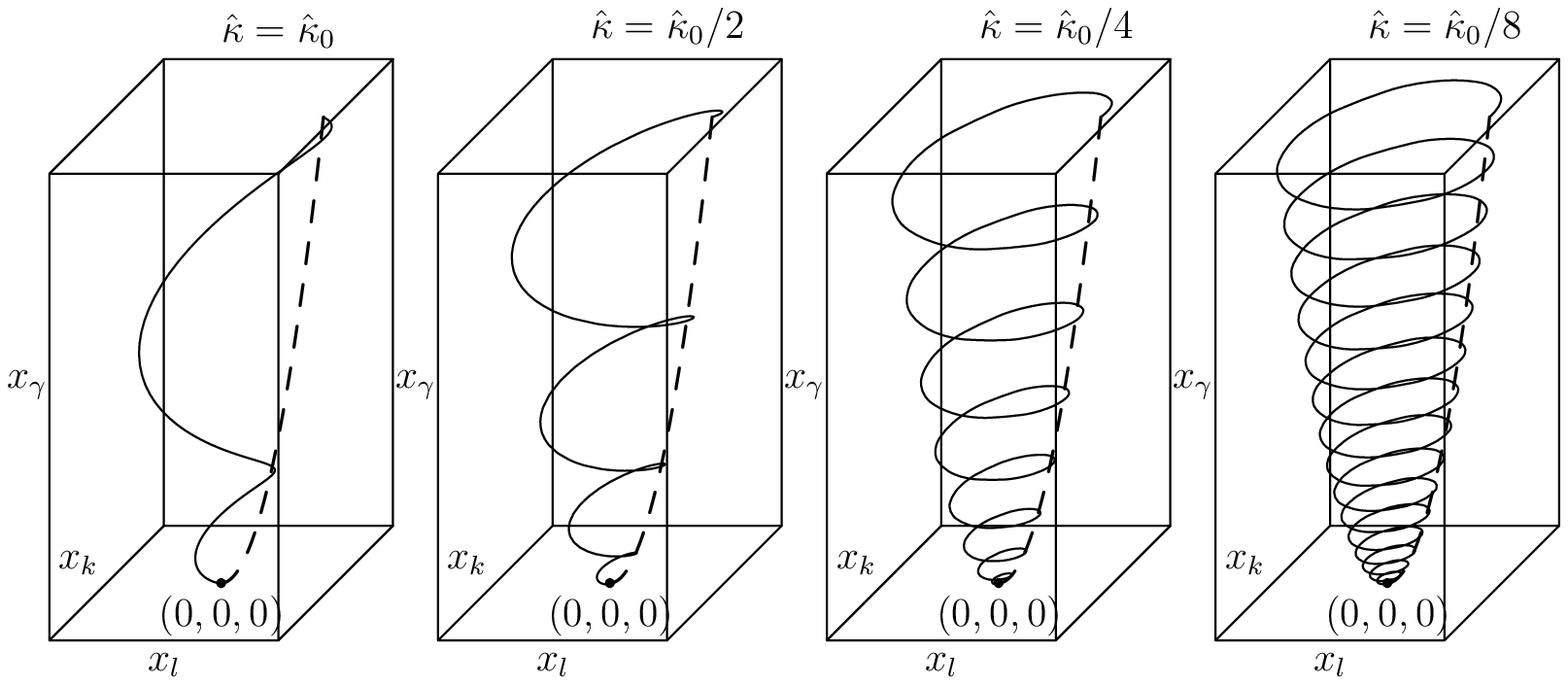}
\end{center}
\caption{The illustration of particle trajectories for different
values of parameter ${\hat \kappa}$ $({\hat \kappa}_{0}=30)$ with
nonzero value of $m$, $F_\gamma$ and  $v_{l0}$ only. The dashed line
corresponds to  undeformed case $({\hat \kappa}=\infty)$, and  the
time parameter runs from 0 to ${\frac{4\pi {\hat
\kappa}_{0}}{F_\gamma}}$.} \label{fig.1}
\end{figure}
Their shape indicates that   particle moves in $\gamma$-direction
along  vortex line with period $T=\frac{2\pi{\hat
\kappa}}{F_\gamma}$. Using the solutions (\ref{sso0})-(\ref{ssol})
one can also check that distance between two neighboring rolls of
vortex in  (k,l)-plane is given by

\begin{equation}
\Delta r = 2\pi\sqrt{\left(-x_{k0}+\frac{v_{l0}{\hat
\kappa}}{F_\gamma}-\frac{F_k{\hat \kappa}^2}{F_\gamma^2m}
 \right)^2+\left(x_{l0}+\frac{v_{k0}{\hat
\kappa}}{F_\gamma}+\frac{F_l{\hat
\kappa}^2}{F_\gamma^2m}\right)^2}\;.
\end{equation}
\\
Besides, it should be noted that for both components $F_k$, $F_l$
and all initial constants equal zero, the distance $\Delta r$
vanishes, i.e.  particle moves along  straight line in
$\gamma$-direction with
 constant acceleration $a_\gamma = \frac{F_\gamma}{2m}$.  Of course,
 for parameter ${\hat \kappa} = \infty$,
the solutions (\ref{sso0})-(\ref{ssol}) become undeformed and
describe the motion of classical particle in constant force
$\vec{F}$.

\section{{{Quadratic noncommutativity}}}

Let us now consider the most complicated type of noncommutativity,
i.e. the quadratic  deformation of classical space \cite{kowclas}
(see also \cite{3c})
\begin{equation}
\{\,x_k,x_\gamma\,\} = \frac{1}{\bar \kappa}tx_l\;\;\;,\;\;\;
\{\,x_l,x_\gamma\,\} = -\frac{1}{\bar
\kappa}tx_k\;\;\;,\;\;\;\{\,x_k,x_l\,\} = 0\;,\label{qajj}
\end{equation}
with   dimensionfull  parameter ${\bar \kappa}$ ($\left[\,{\bar
\kappa}\,\right] = {\rm N}\cdot {\rm s}^2$);  indices $k$, $l$,
$\gamma$ are different and fixed. Its relativistic
counterpart has been proposed in \cite{lie2} and \cite{slawa}. \\
The remaining phase space relations are given by
\begin{eqnarray}
&&~~\{\,p_k,x_\gamma\,\} = \frac{1}{\bar \kappa}tp_l\;,\;
 \{\,p_l,x_\gamma\,\} = -\frac{1}{\bar \kappa}
tp_k\;,\;\{\,x_i,p_j\,\}
=\delta_{ij}\;,\cr
&&~~\cr
&&~~~~~~~~~~~~~~~~~~~\{\,x_\gamma,p_\gamma\,\}=1\;,\;\{\,p_a,p_b\,\} =0
 \;,\nonumber
\end{eqnarray}
with $i,j \ne \gamma$ and $a,b=1,2,3$. They satisfy the Jacobi identity together with (\ref{qajj}). \\
One can check that  corresponding  equations of motion  take the
form

\begin{equation}
\dot{x}_{k} \;=\;  \frac{p_{k}}{m} \;+\; tx_{l}\;\frac{1}{\bar
\kappa}\frac{\partial V}{\partial x_\gamma}\;\;\;,\;\;\; \dot{p}_{k}
\;=\;  tp_{l}\;\frac{1}{\bar \kappa}\frac{\partial V}{\partial
x_\gamma}\;-\;\frac{\partial V}{\partial x_{k}}\;,~~
\label{zazadrug1}
\end{equation}

\begin{equation}
 \dot{x}_{l} \;=\;\frac{p_{l}}{m} \;-\; tx_{k}\;\frac{1}{\bar
\kappa}\frac{\partial V}{\partial x_\gamma}\;\;\;,\;\;\;
 \dot{p}_{l} \;=\; -\frac{\partial V}{\partial
x_{l}} \;-\; tp_{k}\;\frac{1}{\bar \kappa}\frac{\partial V}{\partial
x_\gamma}\;, \label{zazadrug2}
\end{equation}
\\
in  $k$, $l$-directions, and

\begin{eqnarray}
\dot{x}_{\gamma} =  \frac{p_{\gamma}}{m} - tx_{l}\,\frac{1}{\bar
\kappa}\frac{\partial V}{\partial x_k}  + tx_{k}\,\frac{1}{\bar
\kappa}\frac{\partial V}{\partial x_l}
\;\;\;,\;\;\;\dot{p}_{\gamma} = -\frac{\partial V}{\partial
x_{\gamma}}\;,
\end{eqnarray}
\\
in  $\gamma$-direction. \\
By direct calculation one can also find  the corresponding Newton
equations, which look as follows
\begin{eqnarray}
m\ddot{x}_k  &=&  -\frac{\partial V}{\partial
x_k}+tx_l\,\frac{m}{\bar \kappa}\frac{d}{dt}\left(\frac{\partial
V}{\partial x_\gamma}\right) +(2t\dot{x}_l + x_l)\,\frac{m}{\bar
\kappa}\frac{\partial V}{\partial x_\gamma}+\nonumber \\
&&~~~~~~~~~~~~~~~~~~~~~~+ t^2 x_k\,\frac{1}{m{\bar
\kappa}^2}\left(\frac{\partial V}{\partial x_\gamma}\right)^2\;,
\label{xxlie2newton0}\\
&~~&~\cr m\ddot{x}_l  &=&  -\frac{\partial V}{\partial
x_l}-tx_k\,\frac{m}{\bar \kappa}\frac{d}{dt}\left(\frac{\partial
V}{\partial x_\gamma}\right) -(2t\dot{x}_k+x_k)\,\frac{m}{\bar
\kappa}\frac{\partial V}{\partial x_\gamma} +\nonumber \\
&&~~~~~~~~~~~~~~~~~~~~~~+t^2 x_l\,\frac{1}{m{\bar
\kappa}^2}\left(\frac{\partial V}{\partial x_\gamma}\right)^2\;,
\label{xxlie2newton00}\\
&~~&~\cr m\ddot{x}_\gamma  &=&  -\frac{\partial V}{\partial
x_\gamma} -tx_l\,\frac{m}{\bar
\kappa}\frac{d}{dt}\left(\frac{\partial V}{\partial x_k}\right)
-(t\dot{x}_l +x_l)\,\frac{m}{\bar \kappa}\frac{\partial
V}{\partial x_k}+\nonumber\\
&~~&~\cr &&~~~~~~~~~~~~~+
 tx_k\,\frac{m}{\bar \kappa}\frac{d}{dt}\left(\frac{\partial V}{\partial
 x_l}\right)+
(t\dot{x}_k+ x_k)\,\frac{m}{\bar \kappa}\frac{\partial V}{\partial
x_l}
 \;.
\label{xxlie2newton1}
\end{eqnarray}\\
Obviously, for the potential (\ref{potential}) the set of equations
of motion takes the form

\begin{equation}
~~~~~~~~~~~\left\{\begin{array}{rcl} m\ddot{x}_k  &=&  F_k
-{\dfrac{m}{\bar \kappa}}F_\gamma\left(x_l - \dfrac{1}{\bar
\kappa}F_\gamma t^2x_k\right) -\dfrac{2m}{{\bar \kappa}}F_\gamma
t\dot{x}_l \;\\
&~~&~\cr m\ddot{x}_l  &=&  F_l +{\dfrac{m}{\bar \kappa}}F_\gamma
\left(x_k + \dfrac{1}{\bar \kappa}F_\gamma t^2x_l
\right)+{\dfrac{2m}{\bar \kappa}}F_\gamma
t\dot{x}_k \;\\
&~~&~\cr m\ddot{x}_\gamma  &=&  F_\gamma +{\dfrac{m}{\bar
\kappa}}F_k\left(t\dot{x}_l +x_l\right)- {\dfrac{m}{\bar
\kappa}}F_l\left(t\dot{x}_k+ x_k\right)
 \;.
\label{xxxlie2newton1}\end{array}\right.
\end{equation}
\\
We see that as in the Lie-algebraic case, the quadratic
noncommutativity generates the velocity and position-dependent
forces, but this time,  with time dependent coefficients
linear and quadratic in time $t$.\\
By direct calculation we  get the solutions
\begin{eqnarray}
{x}_k(t)  &=& A_k(t)\cos\left(\frac{F_\gamma t^2}{2{\bar
\kappa}}\right)-
A_l(t)\sin\left(\frac{F_\gamma t^2}{2{\bar \kappa}}\right)\;,\label{zadrugizm190}\\
&~~&~\cr {x}_l(t)  &=& A_k(t)\sin\left(\frac{F_\gamma t^2}{2{\bar
\kappa}}\right)+
A_l(t)\cos\left(\frac{F_\gamma t^2}{2{\bar \kappa}}\right)\;,\label{zadrugizm191}\\
&~~&~\cr {x}_\gamma (t)  &=& \frac{F_\gamma}{2m}t^2 + v_{\gamma 0}t
+x_{\gamma 0} + \frac{1}{{\bar \kappa}}\int_{0}^{t}\left(z
x_{l}(z)F_k - zx_k(z)F_l\right)dz\;,\label{zadrugizm192}
\end{eqnarray}
\\
where  the coefficients  $A_k(t)$ and $A_l(t)$ are given by

\begin{eqnarray}
A_{k}(t)  &=& x_{k0} + v_{k0}t + \frac{F_k{\bar
\kappa}}{mF_\gamma}\left[\,\pi t\sqrt{\frac{F_\gamma}{\pi{\bar
\kappa}}}\,C\left(t\sqrt{\frac{F_\gamma}{\pi{\bar
\kappa}}}\right)-\sin\left(\frac{F_\gamma t^2}{2{\bar
\kappa}}\right) \,\right] + \nonumber \\&~~&~\cr &+&\frac{F_l{\bar
\kappa}}{mF_\gamma}\left[\,\cos\left( \frac{F_\gamma t^2}{2{\bar
\kappa}}\right)-1 + \pi t\sqrt{\frac{F_\gamma}{\pi{\bar
\kappa}}}\,S\left(t\sqrt{\frac{F_\gamma}{\pi{\bar
\kappa}}}\right)\,\right]\;,\label{nadnarod}\\
&~~&~\cr A_{l}(t)  &=& x_{l0} + v_{l0}t + \frac{F_l{\bar
\kappa}}{mF_\gamma}\left[\,\pi t\sqrt{\frac{F_\gamma}{\pi{\bar
\kappa}}}\,C\left(t\sqrt{\frac{F_\gamma}{\pi{\bar
\kappa}}}\right)-\sin\left(\frac{F_\gamma t^2}{2{\bar
\kappa}}\right) \,\right] + \nonumber \\&~~&~\cr &-&\frac{F_k{\bar
\kappa}}{mF_\gamma}\left[\,\cos\left( \frac{F_\gamma t^2}{2{\bar
\kappa}}\right)-1 + \pi t\sqrt{\frac{F_\gamma}{\pi{\bar
\kappa}}}\,S\left(t\sqrt{\frac{F_\gamma}{\pi{\bar
\kappa}}}\right)\,\right]\;,\label{nadnarod1}
\end{eqnarray}
\\
and the functions $C(z)$, $S(z)$ are defined as follows
\begin{equation}
C(z) =\int_{0}^{z}\cos\left(\frac{\pi
t^2}{2}\right)dt\;\;\;\;\;,\;\;\;\;\;S(z)
=\int_{0}^{z}\sin\left(\frac{\pi t^2}{2}\right)dt\;.
\end{equation}
The corresponding trajectories  for different values of parameter
${\bar \kappa}$ are illustrated on Figure 2. 
Of course, for  deformation parameter approaching  infinity, the
above solution becomes undeformed and describes the classical
particle in a field of constant force.

\begin{figure}[!h]
\begin{center}
\includegraphics[width=14cm]{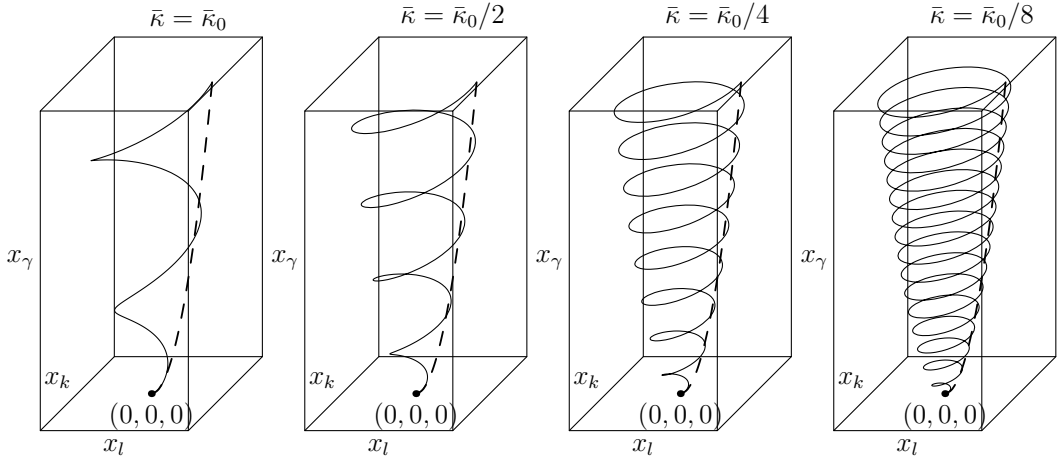}
\end{center}
\caption{The illustration of particle trajectories for different
values of parameter ${\bar \kappa}$ $({\bar \kappa}_{0}=30)$ with
nonzero value of $m$, $F_\gamma$ and  $v_{l0}$ only. The dashed line
corresponds to  undeformed case $({\bar \kappa}=\infty)$, and  the
time parameter  runs from 0 to $2\sqrt{\frac{2\pi {\bar
\kappa}_{0}}{F_\gamma}}$.} \label{fig.2}
\end{figure}

\section{{{Final remarks}}}

In this article we investigate  properties of  simple
classical system 
in the presence of three known noncommutative manifolds: canonical,
Lie-algebraic and quadratic  space-times. We indicate that there are
no dynamical effects for soft type of deformation, while for the
Lie-algebraic and quadratic noncommutativities there appear
additional velocity
and position-dependent  forces. 
The solutions of corresponding Newton equations are provided  and
analyzed in this paper.

The present studies can be extended in various way.
First of all, one can 
consider  more complicated
  system  like  the particle in a presence of well-known   harmonic oscillator potential
 - see e.g. \cite{romero}. Unfortunately, due to the complicated form
of the Newton equations (\ref{lie2newton0})-(\ref{lie2newton1}) and
(\ref{xxlie2newton0})-(\ref{xxlie2newton1}) such  question appears
highly nontrivial and is postponed for subsequent
investigations mainly. 
It is also  interesting to consider basic  quantum systems in
noncommutative space as for example
  the particle in a hole potential, and find their spectra in the presence of all considered
deformations (see e.g. \cite{lodzianieosc}). Finally, one can extend
the presented studies to the case of deformed relativistic particle
at the classical and quantum level (see \cite{deri}, \cite{mech1},
\cite{mechju}). The investigations in these directions already
started and  are  in progress.

\section*{Acknowledgments}
The authors would like to thank J. Lukierski, P. Garbaczewski,  A.
Frydryszak, Z. Jaskolski, J. Kowalski-Glikman and M. Woronowicz
for valuable discussions.\\
This paper has been financially supported by Polish funds for
scientific research (2008-10) in the framework of a research
project.


\end{document}